\documentclass{IAU-TrA}
\usepackage{graphicx}

\title[Atomic and Molecular Data]     %% header right hand page %%
{}

\author[DIVISION~B / COMMISSION 14]   %% header left hand page %%
{}

\pubyear{2015}
\volume{Volume 29A}
\pagerange{001--008}
\jname{Transactions IAU, Volume 29A}
\editors{Thierry Montmerle, ed.}
\begin{document}

\maketitle

{\bf

\large
\begin{tabbing}
\hspace*{65mm}       \=                                              \kill
COMMISSION~14         \> ATOMIC AND                            \\
                     \> MOLECULAR DATA                            \\[0.5ex]
                     \> {\small\it DONNEES ATOMIQUE ET MOLECULAIRES}                 \\
\end{tabbing}

\normalsize

\begin{tabbing}
\hspace*{65mm}       \=                                              \kill
PRESIDENT            \> Lyudmila I. Mashonkina                              \\
VICE-PRESIDENT       \> Farid Salama                          \\
PAST PRESIDENT       \> Glenn M. Wahlgren                      \\
ORGANIZING COMMITTEE \> France Allard, Paul Barklem,              \\ 
                     \> Peter Beiersdorfer, Helen Fraser,         \\
                     \> Gillian Nave, Hampus Nilsson          \\
\end{tabbing}

\noindent
COMMISSION~14 WORKING GROUPS
\smallskip

\begin{tabbing}
\hspace*{65mm}      \=                                               \kill
Div.~B / Commission~14 WG \>  Atomic Data                  \\
Div.~B / Commission~14 WG \>  Collision Processes                       \\
Div.~B / Commission~14 WG \>  Molecular Data       \\
Div.~B / Commission~14 WG \>  Solids and Their Surfaces                    \\

\end{tabbing}

\bigskip

\noindent
TRIENNIAL REPORT 2012-2015
}

\firstsection % if your document starts with a section,

\section{Introduction}

The main purpose of Commission~14 is to foster interactions between the astronomical 
community and those conducting research on atoms, molecules, and solid state particles
 to provide data vital to reducing and analysing  
astronomical observations and performing theoretical investigations. 

Commission~14 supports a website at {\tt http://www.inasan.ru/iau14/}. It informs the astronomical 
community of the meetings of interest, provides links to the relevant databases, contains Triennial Reports of Commission~14 and Working Groups (WGs) for the past triennia, and includes the history of the Commission~14 officers, Organizing Committee (OC), and Working Groups.

\section{Historical background}

Commission~14 has been operating for almost a century. It was one of the 32 Standing IAU Commissions (\cite{Blaauw1994}) and established in 1919, with the name Wavelength Standards and Spectral Tables for the Sun. The first President was C.~E.~St.~John.  
 In 1961 at the IAU General Assembly (GA) XI in Berkeley it was renamed as Commission~14 Fundamental Spectroscopic Data and in 1979 at the IAU GA XVII in Montreal as Commission~14 Atomic and Molecular Data. 

Before 2003 in recognition of its special interdisciplinary character, Commission~14 was linked directly to the IAU Executive Committee. It was included in a newly created Division~XII at the IAU GA XXV in Sydney, July 2003. In 2012 the IAU GA XXVIII in Beijing approved the new Divisional restructuring, and Commission~14 became a part of Division~B.

In the past twenty years Commission~14 was led by Presidents W.~H.~Parkinson (1994-1997), F.~Rostas (1997-2000), P.~L.~Smith (2000-2003), S.~Johansson (2003-2006), S.~R.~Federman (2006-2009), G.~M.~Wahlgren (2009-2012), and L.~Mashonkina (2012-2015). 
In different triennia, the Organising Committee included seven to twelve members, and the usual practice was for a member to serve for six years. 

Up to eight WGs operated successfully within Commission~14 during each triennium. They were chaired by recognised experts, i.e. WG Atomic Spectra and Wavelengths (S.~Johansson, G.~Nave), WG Atomic Transition Probabilities (J.~Fuhr, G.~M.~Wahlgren, W.~L.~Wiese), WG Atomic Data (J.~Fuhr, G.~Nave, G.~M.~Wahlgren), WG Collision Processes (P.~S.~Barklem, M.~S.~Dimitrijevic, G.~Peach, D.~R.~Schultz, P.~C.~Stancil), WG Line Broadening (G.~Peach, C.~Stehle), WG Molecular Structure (E.~F.~Van~Dishoeck), WG Molecular Reactions on Solid Surfaces (W.~Schutte), WG Molecular Data (P.~Bernath, J.~Black, S.~Federman, H.~M\"{u}ller), WG Gas Phase Reactions (T.~J.~Millar), and WG Solids and Their Surfaces (T.~Henning, H.~Linnartz, G.~Vidali). 

\section{Activity}

One way that the Commission accomplishes its goal is through triennial 
compilations on recent relevant research in atomic, molecular and solid state physics, 
 and chemical analysis, and their application to astronomy. The most recent compilations appear in the accompanying set of Commission~14 WG Triennial Reports, which were produced by members of the 
WGs and the OC (\cite{2015mol}; \cite{2015atom}; \cite{2015coll}; \cite{2015solids}). 
Triennial Reports for the past triennia, starting from 2000-2003, 
were published in Transactions IAU (\cite{2003c14}; \cite{2007c14}; \cite{2009c14}; \cite{2012c14}; \cite{2012mol}; \cite{2012atom}; \cite{2012coll}; \cite{2012solids}).

%\section{Sponsored meetings within the past triennium}

During 2012-2015, members of Commission~14 have also been active in organizing and participating in meetings of various types. The following meetings were sponsored by Commission~14.

\begin{itemize}
\item IAU Symposium No. 297 'The Diffuse Interstellar Bands', May 2013, Noordwijkerhout, The Netherlands
\item International conference 'Putting A Stars into Context: Evolution, Environment, and Related Stars', June 2013, Moscow, Russia
\item IAU GA XXIX Focus Meeting 12 (FM12) 'Bridging Laboratory Astrophysics and Astronomy: From Provider to User', August 3-5, 2015, Honolulu, United States
\end{itemize}

The latter meeting was organised by Commission~14 together with the American Astronomical Society Laboratory Astrophysics Division (AAS LAD). Farid Salama, Lyudmila Mashonkina, and Steven Federman co-chaired the SOC. FM12 involved atomic and molecular data, plasma physics, nuclear physics, and particle physics and their application to various fields, such as interplanetary, interstellar, and intergalactic matter, planetary and stellar atmospheres. 

Through these meetings the
astronomical community can communicate their needs to data producers, while data producers
provide the results of their studies. Input from the astronomical community is critical to 
maintaining the vitality of the data producing community, fostering collaboration on proposals and projects that can lead to funding opportunities for data producers. 
We are witnessing a rapid growth in quantity and quality of astronomical measurements that is driven by a combination of new and larger telescopes equipped with more sensitive detectors, and with capabilities to acquire high spectral and spatial resolution data at wavelengths spanning the X-ray, ultraviolet, infrared, sub-millimeter and very long radio wavelength regimes.   
Interpreting these superb astronomical observations requires an understanding of the fundamental properties and processes of atoms and particles, molecules, ions and solids to an unprecedented precision.  
This need has attracted the attention and the interest of laboratory and theoretical scientists from different disciplines who have brought new or improved laboratory techniques, large-scale theoretical calculations and simulations to astronomy.  Modern computers also allow detailed modeling of astrophysical phenomena, with the possibility to include chemical reactions, atomic, molecular, plasma, nuclear and particle processes and their coupling to astrophysical environments. 

%\newpage

\section{Looking forward to the next triennial period}

Commission~14 terminated at the Honolulu General Assembly. However, the field covered by Commission~14 remains with the IAU. By call of the IAU, the Commission~14 OC initiated a proposal for a new Commission
on Laboratory Astrophysics that would be a natural evolution of Commission~14. The Vice-President of Commission~14 Farid Salama, with the active help
of the co-proposers Paul Barklem, Helen Fraser, Thomas Henning, and Gianfranco Vidali, 
was leading this effort. The proposed Commission was accepted in April, 2015 as the IAU Commission B5 Laboratory Astrophysics, with President Farid Salama. Helen Fraser was elected by the commission membership at large as Vice-President and Harold Linnartz and Gianfranco Vidali as OC members, in addition to the ex-officio ones. 

The goal of the new Laboratory Astrophysics Commission is better understanding of the Universe through the promotion of fundamental theoretical and experimental research into the underlying processes that drive the cosmos. To achieve this goal, the Commission intends to facilitate interactions between the international astronomical community (user community) and the experimentalists and theorists who provide the necessary atomic, molecular, solid-state, nuclear and particle astrophysics data. The Commission will adopt a strategy to promote the field of laboratory astrophysics, particularly with reference to ground-based and space-born astronomy missions. To increase the visibility of this initiative the two new mailing lists, namely the American LAD list and the European labastro list, have been initiated.

%%\section{Closing remarks}

\vspace{3mm}
 
{\hfill Lyudmila Mashonkina}

{\hfill {\it President of the Commission}}

\end{document}